\documentclass{optica-article}

\journal{opticajournal}

\newcommand\nnfootnote[1]{%
  \begin{NoHyper}
  \renewcommand\thefootnote{}\footnote{$\dagger$ These authors contributed equally to this work.}%
  \addtocounter{footnote}{-1}%
  \end{NoHyper}
}

\begin{document}

\title{Reciprocity-assisted ghost imaging through dynamic random media}
\author{Edward Tananyan\authormark{1,$\dagger$}, Ohad Lib\authormark{1,$\dagger$},Michal Zimmerman\authormark{1,$\dagger$}, and Yaron Bromberg\authormark{1,*}}

\address{\authormark{1} Racah Institute of Physics, The Hebrew University of Jerusalem, Jerusalem, 91904 Israel}

\email{\authormark{*}yaron.bromberg@mail.huji.ac.il} 
\address{\authormark{$\dagger$}These authors have contributed equally to this work}
\begin{abstract}
Ghost imaging enables the imaging of an object using intensity correlations between a single-pixel detector placed behind the object and a camera that records light that did not interact with the object. The object and the camera are often placed at conjugate planes to ensure correlated illumination patterns. Here, we show how the combined effect of optical reciprocity and the memory effect in a random medium gives rise to correlations between two beams that traverse the random medium in opposite directions. Using these correlations, we demonstrate a ghost imaging scheme in which the object and camera are placed at opposite ends of the random medium and illuminated by counter-propagating beams that can potentially be emitted by two different sources. 
\end{abstract}


Ghost imaging is an imaging technique in which an image of an object can be reconstructed, even though the light that interacted with the object is detected by a single-pixel detector. The spatial information of the object is obtained by correlating the intensities measured by the single-pixel detector with the measurements from a multi-pixel detector, such as a camera\cite{erkmen2010ghost}. The camera probes the intensity pattern of a reference beam that is spatially correlated with the beam illuminating the object. Early demonstrations of ghost imaging relied on quantum correlations between pairs of spatially entangled photons that formed the object and reference beams\cite{pittman1995optical,strekalov1995observation}. However, it was later realized that classical correlations between two identical copies of a spatially inhomogeneous beam are sufficient for obtaining a ghost image \cite{bennink2002two,gatti2004ghost,ferri2005high,valencia2005two}. To clarify the role of classical and quantum correlations in ghost imaging, researchers replaced the reference beam with the computation of the predetermined illumination pattern in a configuration known as computational ghost imaging\cite{shapiro2008computational, bromberg2009ghost}, a spatially coherent variant of dual photography and single-pixel imaging \cite{sen2013relationship,sen2005dual,duarte2008single}.

Common to all ghost imaging configurations is knowing the intensity pattern that illuminates the object. Therefore, the quality of the reconstructed image is sensitive to misalignments, aberrations, or scattering that deteriorate the mapping between the illumination pattern in the object beam and the reference beam (or the virtual reference beam in computational ghost imaging) \cite{gong2011correlated}. When imaging through random media, the non-scattered (ballistic) component of the beams can still be utilized for ghost imaging \cite{bina2013backscattering,xu2015ghost}. However, in the strong scattering regime, the ballistic term vanishes, and the reference and sample beams become presumably uncorrelated. Nevertheless, when using classical coherent light, the scattered speckle patterns carry information on the correlation between the beams, which can be retrieved using phase retrieval algorithms \cite{yuan2022unsighted,zhang2024imaging}. Alternatively, one can use the speckle pattern generated by the random media itself to illuminate an object placed behind it and use correlations between the backscattered and transmitted light for blind ghost imaging \cite{paniagua2019blind}.

In this work, we explore a different configuration of ghost imaging that utilizes the speckle generated by the random media to illuminate the object. To obtain a correlated speckle pattern at the plane of a camera located on the opposite side of the random media, we use a counter-propagating reference beam. Based on optical reciprocity and the memory effect\cite{freund1988memory,feng1988correlations}, we derive and experimentally measure strong correlations between the speckle patterns created by the counter-propagating beams and demonstrate how these correlations can be used for ghost imaging.

\begin{figure}[h!]
\centering\includegraphics[width=7cm]{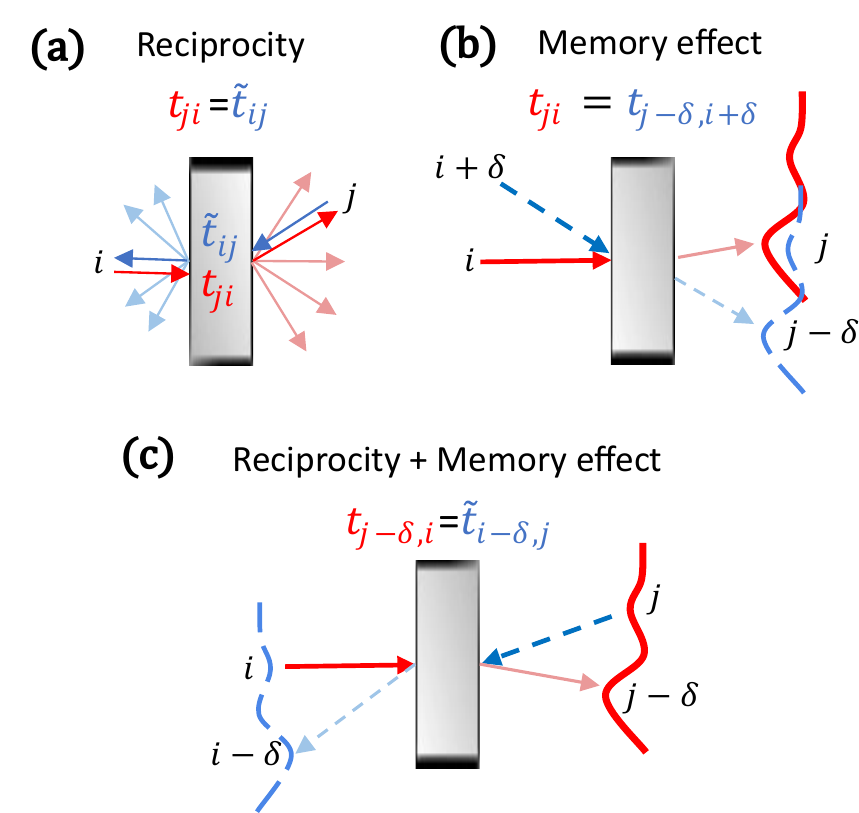}
\caption{\label{fig1} Intensity correlations between counter-propagating beams through a scattering medium. (a) Reciprocity ensures that the intensity at point $j$ resulting from a source at point $i$ is correlated with the intensity at point $i$ resulting from a counter-propagating source at point $j$. $t$ and $\widetilde{t}$ are the left-to-right and right-to-left transmission matrices, respectively. (b) For $\delta$ within the memory effect range of the scattering sample, shifting the source by $\delta$ approximately shifts the output speckle pattern by $-\delta$\cite{freund1988memory,feng1988correlations}. (c) Thanks to both reciprocity and the memory effect, for sources at $j$ and $i$, the intensities at points $i-\delta$ and $j-\delta$ are correlated.}
\end{figure}

We start by explaining the correlations between the speckle patterns created by two counter-propagating beams passing through a random medium. We mark the transmission matrix of the medium for light traveling from left to right by $t$, and from right to left by $\widetilde{t}$. Given sources at point $i$ on the left side of the scattering medium and at point $j$ on the right side, reciprocity yields that the intensity of the resulting speckle patterns at points $i$ and $j$ are correlated (fig.\ref{fig1}a). In terms of the elements of the transmission matrices, reciprocity ensures $t_{ji}=\widetilde{t}_{ij}$. However, since reciprocity only guarantees correlations between two points, it offers limited benefits for imaging. 

To extend the correlation range, one can utilize the memory effect of the scattering medium\cite{freund1988memory,feng1988correlations}. In general, when changing the illumination angle of a beam entering a scattering medium, the output speckle pattern may change dramatically. However, if the change in illumination angle is small enough, within the so-called memory effect range, it was shown that the structure of the output speckle pattern remains largely unchanged, but shifted (fig. \ref{fig1}b). This yields correlations in the transmission matrix $t$, so that for a small enough shift of the illumination angle $\delta$, one can approximately obtain $t_{ji}=t_{j-\delta,i+\delta}$. 

The combined effect of reciprocity and the memory effect thus extends the correlations between the speckle patterns around the reciprocal points to the size of the memory effect range (fig.\ref{fig1}c). This result can be intuitively understood by first considering a source on the left side at point $i-\delta$ and on the right side at point $j$. From reciprocity, the intensities at points $i-\delta$ and $j$ are correlated. Within the memory effect range, shifting the source on the left side from $i-\delta$ to $i$ shifts the speckle pattern on the right side by $-\delta$. This means that the intensity observed at point $j$ is now shifted and observed at point $j-\delta$. Therefore, we can conclude that the intensities at points $i-\delta$ and $j-\delta$ are correlated for sources at points $i$ and $j$, for $\delta$ within the memory effect range. In terms of the transmission matrix elements, we obtain $t_{j,i-\delta}=\widetilde{t}_{i-\delta,j}$ from reciprocity, and $t_{j,i-\delta}=t_{j-\delta,i}$ from the memory effect, yielding the desired correlations $t_{j-\delta,i}=\widetilde{t}_{i-\delta,j}$. We note that these correlations are related to coherent backscattering when considering the reflection matrix rather than the transmission matrix of the scattering medium\cite{freund1990surface}.

\begin{figure}[h!]
\centering\includegraphics[width=7cm]{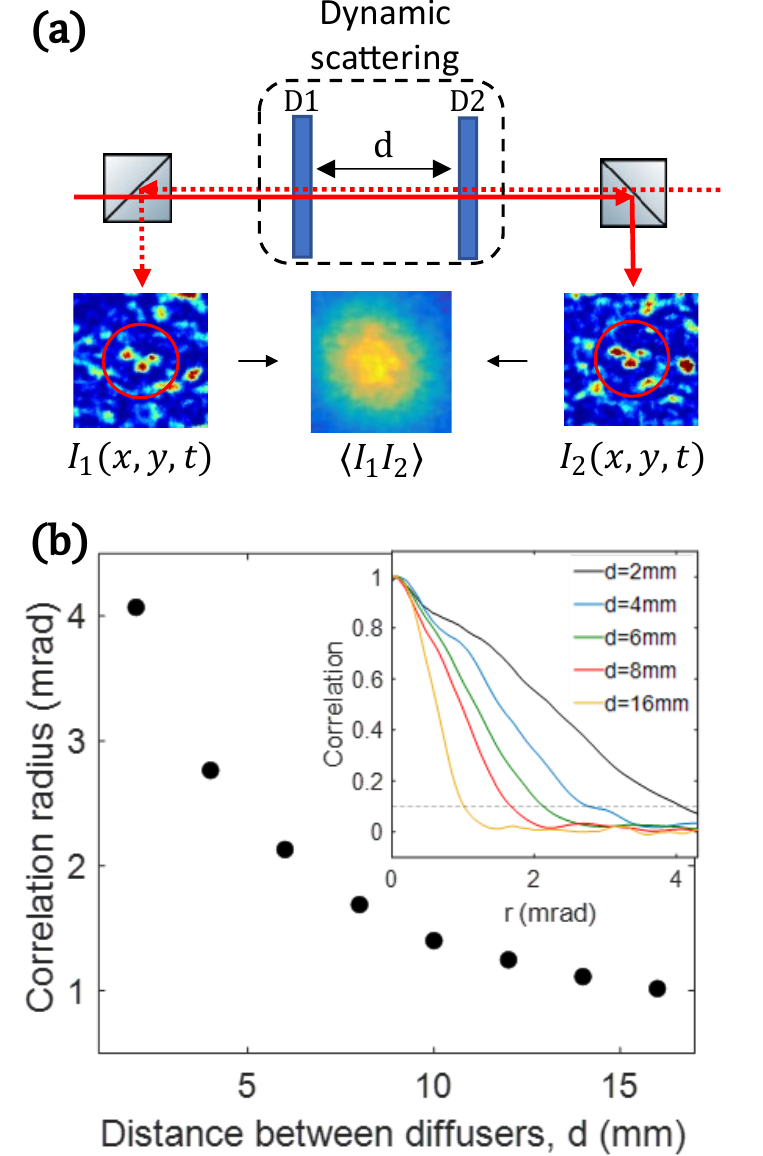}
\caption{\label{fig2} Correlations between speckle patterns of counter-propagating beams. (a) Two collimated counter-propagating beams (solid and dashed arrows) are scattered by a pair of thin diffusers separated by a distance $d$. The intensities $I_1(x,y,t),I_2(x,y,t)$ of the output speckle patterns at the far-field exhibit correlations within a finite range. (b) Intensity correlations as a function of the distance $r$ from the reciprocal points (inset) for different spacing $d$ between the diffusers. The effective correlation radius, defined as the radius at which the correlation decreases to $0.1$, decreases as the diffuser spacing increases.}
\end{figure}

To measure these correlations experimentally, we illuminate a scattering medium with two counter-propagating beams and record the two output speckle patterns, $I_1(x,y,t),I_2(x,y,t)$ using a camera (fig. \ref{fig2}a, supplementary information). The scattering medium consists of two thin diffusers separated by a variable distance $d$, enabling us to control the memory effect range in the experiment. The thin diffusers are rotated to change between different disorder realizations, and the correlations between the speckle patterns are obtained by averaging $I_1(x,y,t)I_2(x,y,t)$ over time (or equivalently, over different disorder realizations). We observe correlations between the speckle patterns created by the counter-propagating beams, which gradually decrease when the distance $r$ from the reciprocal points increases, as expected (fig.\ref{fig2}b, inset). The correlation radius increases when decreasing the distance $d$ between the diffusers, as shown in fig.\ref{fig2}b, due to the increase in the memory effect range.

\begin{figure}[h!]
\centering\includegraphics[width=7cm]{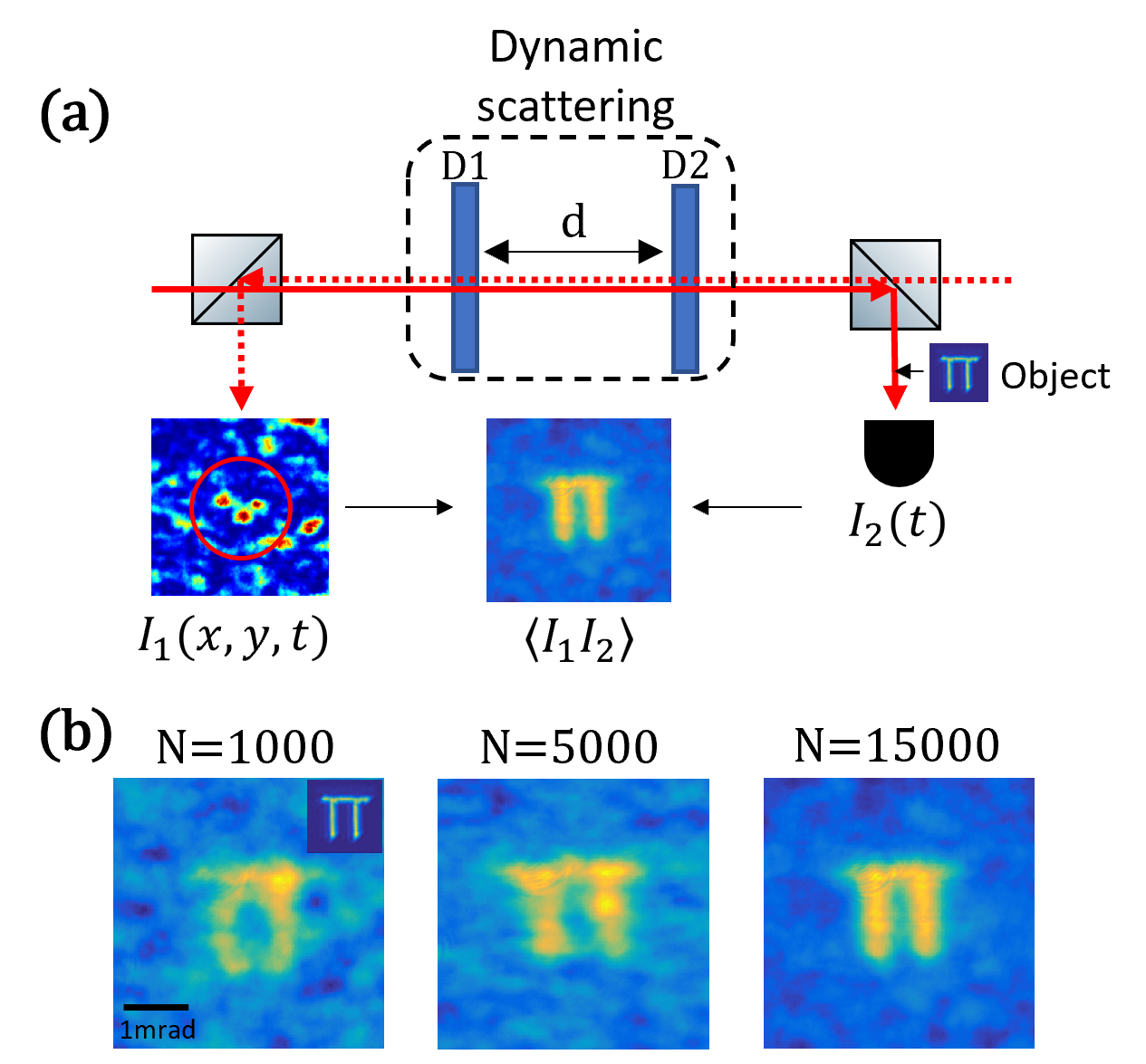}
\caption{\label{fig3} Ghost imaging using counter-propagating beams. (a) Two counter-propagating beams pass through a dynamic random medium (solid and dashed arrows). An object, smaller than the memory effect range, is placed at the path of one of the beams, and the intensity behind it $I_2(t)$ is recorded using a bucket detector without any spatial resolution. By correlating the intensity of the bucket detector with the spatially resolved intensity of the counter-propagating beam that does not pass through the object, an image of the object can be recovered. (b) The quality of the formed image increases with the number of recorded frames, $N$. All images were recorded with $d=2mm$.}
\end{figure}

After establishing the correlations between speckle patterns created by counter-propagating beams, we utilize this effect for ghost imaging. The object we wish to image is located on the right side of the dynamic scattering medium and illuminated by a time-varying speckle pattern. The light that passes through the object is measured using a bucket detector with no spatial resolution, yielding a signal $I_2(t)$ (fig.\ref{fig3}a). The speckle pattern resulting from the counter-propagating beam, which has never interacted with the object, is recorded with a camera on the left side of the scattering medium, $I_1(x,y,t)$. By correlating the signals from the camera and bucket detector and averaging over time, an image of the object is reconstructed (fig.\ref{fig3}b). The image reconstruction quality increases with the number of disorder realizations used.

In conclusion, we have theoretically discussed and experimentally measured intensity correlations between speckle patterns of counter-propagating beams resulting from reciprocity and the memory effect of the scattering medium. We have utilized these speckle correlations to experimentally realize a ghost imaging scheme in which the object and camera are illuminated via two beams counter-propagating through a dynamic random medium. Our ghost imaging scheme can be particularly useful in scenarios where the imaging camera and object are located on different sides of a dynamic scattering medium that exhibits a relatively large memory effect range, such as Earth's turbulent atmosphere\cite{beckers1988increasing}. In the case of atmospheric turbulence, the two counter-propagating beams may even be at different wavelengths, thanks to the low dispersion of air \cite{lib2020real}. Interestingly, as weak correlations are sufficient for ghost imaging, we expect our method to have a relatively wide field of view even when the scattering medium exhibits a narrow memory range. Finally, following the large number of imaging schemes through random media that rely on the memory effect\cite{bertolotti2012non,katz2014non,edrei2016memory,li2021memory}, we expect correlations between counter-propagating beams to potentially facilitate new imaging schemes beyond ghost imaging. 

\begin{backmatter}
\bmsection{Funding}
This research was supported by the Zuckerman STEM Leadership Program and the Israel Science Foundation (grant No. 2497/21).

\bmsection{Acknowledgments}
We thank Ori Katz for fruitful discussions. 

\bmsection{Disclosures}
The authors declare no conflicts of interest.

\bmsection{Data Availability Statement}
Data underlying the results presented in this paper are not publicly available at this time but may be obtained from the authors upon reasonable request.

\bmsection{Supplemental document}
See Supplement 1 for supporting content.
\end{backmatter}

\bibliography{sample}

\def\thefigure{S\arabic{figure}}
\setcounter{figure}{0}

\section*{Supplementary information: experimental setup}

Two counter-propagating beams at wavelength $\lambda=640nm$ pass through a pair of diffusers ($D1,D2$) separated by a variable distance $d$. A portion of each beam is then reflected by a beamsplitter (BS) and imaged using a CCD camera at the Fourier plane (passing through a $f=250mm$ lens). A polarizing beamsplitter (PBS) ensures a good contrast of the speckle patterns. For ghost imaging measurements, an object is inserted in the path of one of the beams, and the corresponding area of the camera is used as a bucket detector by summing the intensity of all relevant pixels. To easily find the correlated spots of the two beams, we replaced the scattering medium with a white piece of paper and found the coherent backscattering peaks of both beams, which mark the effective income angle of the source.

\begin{figure}[h!]
\centering\includegraphics[width=7cm]{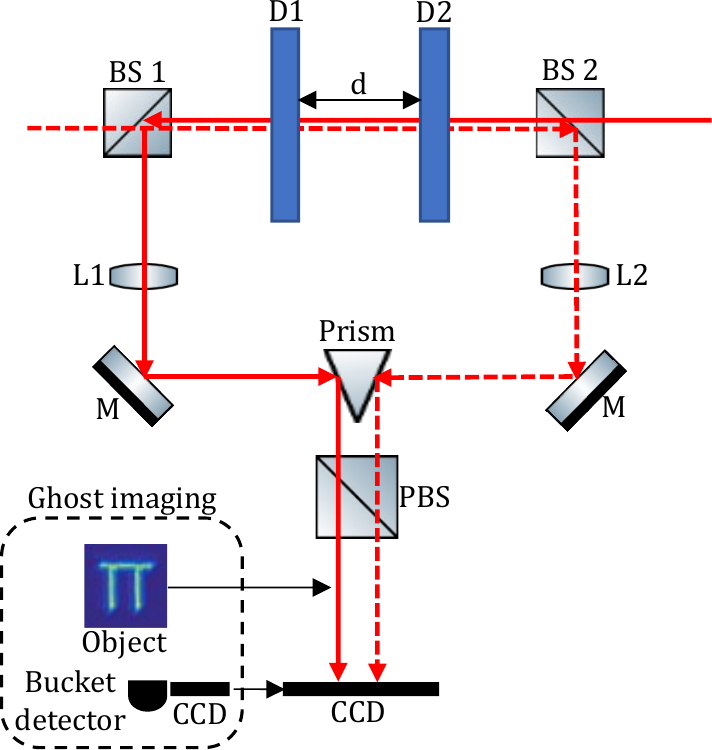}
\caption{\label{figS1} Illustration of the experimental setup. See text for details. BS - beamsplitter, D - diffuser, M - mirror, L - lens, PBS - polarizing beamsplitter.}
\end{figure}

\end{document}